\documentclass[letterpaper]{article}

\usepackage{natbib,alifeconf}
\PassOptionsToPackage{hyphens}{url}\usepackage{hyperref}

\title{Seeking Open-Ended Evolution in Swarm Chemistry II:\\
Analyzing Long-Term Dynamics via Automated Object Harvesting}
\author{Hiroki Sayama\\
\mbox{}\\
Center for Collective Dynamics of Complex Systems, 
Binghamton University, State University of New York\\
Binghamton, NY 13902-6000, USA \quad
sayama@binghamton.edu}

\begin{document}
\maketitle

\begin{abstract}
We studied the long-term dynamics of evolutionary Swarm Chemistry by extending the simulation length ten-fold compared to earlier work and by developing and using a new automated object harvesting method. Both macroscopic dynamics and microscopic object features were characterized and tracked using several measures. Results showed that the evolutionary dynamics tended to settle down into a stable state after the initial transient period, and that the extent of environmental perturbations also affected the evolutionary trends substantially. In the meantime, the automated harvesting method successfully produced a huge collection of spontaneously evolved objects, revealing the system's autonomous creativity at an unprecedented scale.
\end{abstract}

\section{Introduction}

Open-ended evolution (OEE) has been one of the major issues discussed in Artificial Life (ALife) research \citep{bedau1992measurement,bedau1998classification,bedau1999visualizing,bedau2000open}, which is recently regaining significant attention from the ALife community and beyond \citep{soros2014identifying,taylor2016open,stanley2017open}. In the literature, OEE is often characterized by measuring the ongoing activity level of producing novel adaptations in evolving individuals (e.g., genotypes) \citep{bedau1992measurement,bedau1998classification,bedau1999visualizing}. Such characterization implicitly assumes that there are clearly identifiable ``individuals'' as evolving entities, and that the mechanisms that drive evolution, such as variation and selection, are already in place within the evolutionary system under investigation.

Those assumptions mentioned above are not readily applicable, however, to more implicit, emergent evolutionary systems in which ``individuals,'' ``variation,'' and ``selection'' are all observed phenomenologically as higher-level emergent properties that arise from interactions among distributed lower-level components, such as in Artificial Chemistry (AChem) models \citep{dittrich2001artificial,banzhaf2015artificial,schmickl2016life}. Studying OEE in such emergent ALife/AChem models provides ample opportunities of novel research programs, potentially merging OEE with other research on the origins of life and/or self-organization of complex systems.

To explore the possibility of OEE in highly decentralized AChem-based systems, we have been developing and evaluating evolutionary versions of Swarm Chemistry \citep{sayama2009swarm,sayama2011seeking,sayama2011quantifying,sayama2012evolutionary,sayama2018complexity}. In these models, a fixed number of self-propelled particles move and stochastically differentiate into one of the kinetically distinct types specified in the behavioral parameter settings (called ``recipes'') they carry. Their physical contacts (collisions) may cause transfer of the recipe information from one particle to another, depending on a locally defined ``competition function'' that serves as the basic law of ``physics'' of this simulated world. Such recipe transmissions occur at a microscopic, individual particle level, while there is no explicit fitness function that determines  selection criteria for self-organizing macroscopic structures. These models come with no \textit{a priori} given macroscopic descriptors of evolutionary dynamics, and therefore, what are the evolving ``individuals'' and how they are ``adapting'' in the simulated world are ultimately up to the observer's interests and interpretations.

Evolutionary Swarm Chemistry models have demonstrated, under certain experimental conditions, continuous production of novel macroscopic structures for a substantially long period of time \citep{sayama2011seeking,sayama2011quantifying}, especially in two-dimensional space \citep{sayama2012evolutionary}. In the meantime, several important questions remain unanswered, including (1) whether their dynamics exhibit true OEE for indefinitely long duration of time, and (2) how one can detect and extract morphological structures (which may be considered emergent ``individuals'' in this system) automatically from simulation runs. 

In this paper, we aim at addressing the aforementioned two questions by extending the simulation length ten-fold compared to that used in earlier work, and by developing a novel, image processing-based method for automated harvesting of morphological structures from simulation results. In what follows, we describe the outline of the simulation model and the experimental conditions, the newly developed automated object harvesting method, and analytical methods used to characterize the observed evolutionary dynamics. We report results and findings obtained from a dozen of extended simulation runs, especially on how the evolutionary dynamics changed over time and how they were affected by different forms of environmental perturbations.

\section{Methods}

\subsection{Simulation model and settings}

We used the evolutionary Swarm Chemistry model described in \citep{sayama2011seeking}, which was based on Swarm Chemistry \citep{sayama2009swarm}, a computational model of collective behaviors of self-propelled particles
similar to Reynolds' ``Boids''
\cite{reynolds1987}. In Swarm Chemistry, multiple types of particles
with different kinetic behavioral parameters are mixed together. Their
behavioral parameters are represented in a ``recipe,'' a set of sequences of kinetic parameter values each of which describes the strength of a specific behavioral rule such as \textit{cohesion}, \textit{alignment}, \textit{separation} \citep{reynolds1987}, etc. Each parameter value sequence in a recipe represents behaviors of one particle type.

In the evolutionary Swarm Chemistry models, these recipes evolve through their transmissions (possibly with stochastic mutations) among colliding particles. The direction
of recipe transmission is determined by a competition function that takes local information about two particles engaged in a collision and returns which one becomes the source of the recipe transmission. Examples of competition functions include: ``faster'' (the faster particle wins = becomes the source of recipe transmission), ``slower'' (the slower one wins), ``majority'' (the one that is surrounded by more particles of the same type wins), and so on. The
competition function may be varied temporarily as well as spatially to induce
exogenous environmental perturbations to promote continuous evolutionary activities.

In this study, we adopted the ``revised-high'' experimental condition that was identified in \citep{sayama2011quantifying} to be most successful in producing continuous evolutionary dynamics of macroscopic nontrivial structures. This condition uses a revised collision detection method described in \citep{sayama2011quantifying}, high mutation rates, and spatially heterogeneous environmental perturbations. Each simulation was conducted with 10,000 particles moving and interacting within a 5,000$\times$5,000 (in arbitrary unit) continuous two-dimensional space. Boundary conditions were such that the space was toroidal with regard to particles' positions (i.e., they would come out from the opposite edge of the space when they went out) but it was also a bounded square with regard to particles' perception (i.e., they would not sense other particles' presence across the boundaries), following the model settings used in \citep{sayama2011seeking}. Out of the 10,000 particles, 100 were initialized with randomly generated recipes, while the rest were initially inactive. The primary competition function used to decide the direction of recipe transmission was ``majority (relative)'' (see \citep{sayama2011seeking} for details). The simulator code was written in Java and is available from the author's website\footnote{\url{http://bingweb.binghamton.edu/~sayama/SwarmChemistry/}}.

In the present study, each simulation was run for 300,000 time steps, ten times longer than the typical simulation length used in earlier work. Snapshots of simulations were saved in the form of a 4,000$\times$4,000-pixel high-resolution bitmap image in every 1,000 time steps, which were analyzed by a separate process described in the next subsection. In those snapshots, particles' behavioral properties were partly visualized in their colors by mapping the strengths of their \textit{cohesion}, \textit{alignment} and \textit{separation} tendencies to (R, G, B) values.

Like in \citep{sayama2011seeking,sayama2011quantifying}, exogenous environmental perturbations were periodically introduced in every 2,000 time steps onto either left or right half of the space (randomly selected each time) by changing the competition function within the selected half to either ``faster'' or ``slower'' (again, randomly selected each time). In earlier work, each perturbation lasted only for 50 time steps, while in the present study we tested two lengths of environmental perturbations to evaluate the sensitivity of results and the adaptability of evolving structures: (i) 50 time steps (i.e., 1,950 normal steps interjected by 50 steps with perturbation), and (ii) 500 time steps (i.e., 1,500 normal steps interjected by 500 steps with perturbation), which are called ``short'' and ``long'' perturbations, respectively, in the following sections.

\subsection{Automated object harvesting}

The bitmap images of simulation snapshots generated above were processed concurrently by another process for automated object harvesting. In this work, objects are defined as spatially contiguous morphological structures in which particles were clustered to show some kind of organization. Such objects were detected and extracted in the following image processing steps (also see Fig.~\ref{image-processing}):
\begin{figure}[tp]
\centering
\includegraphics[width=\columnwidth]{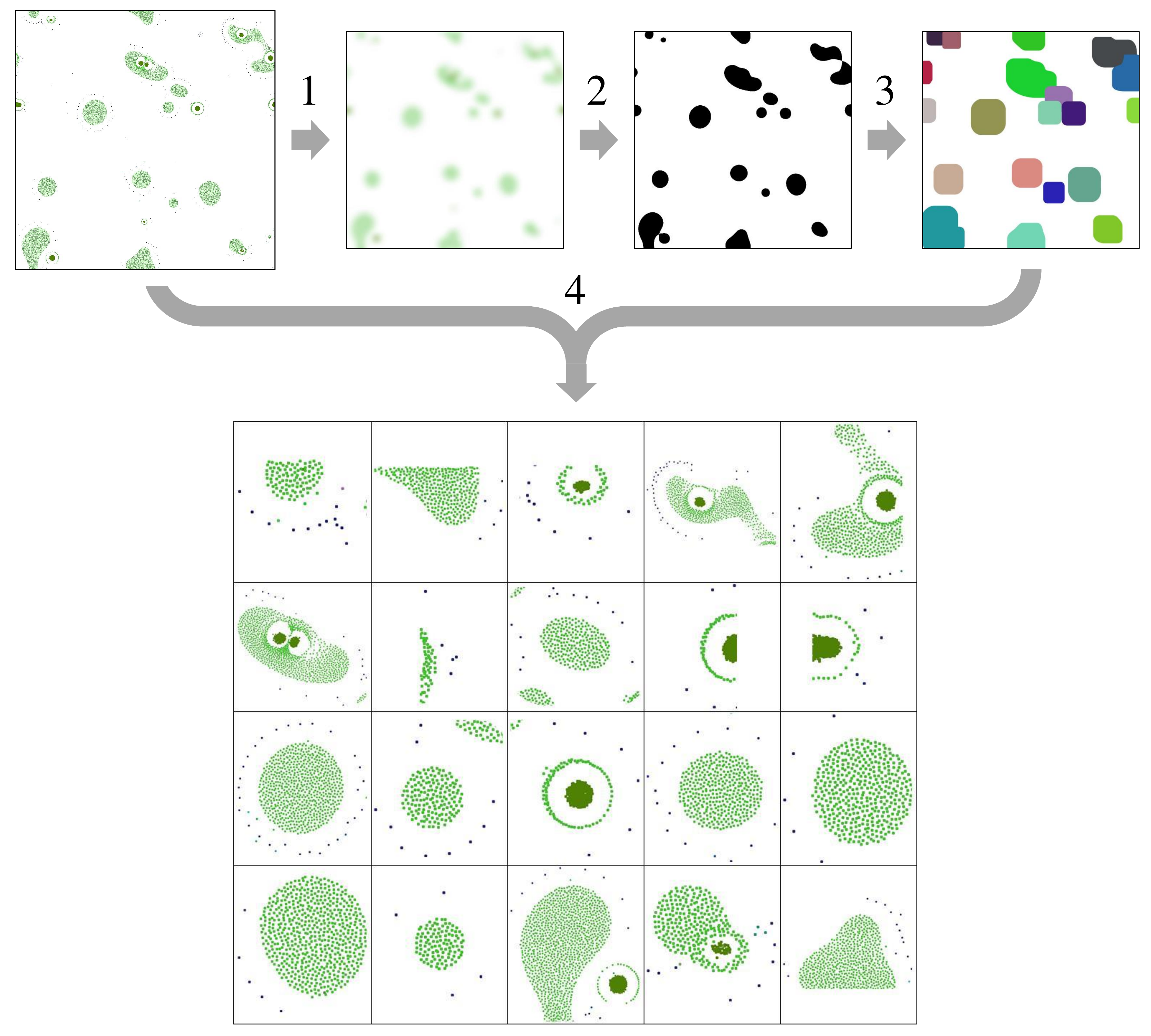}
\caption{Automated object harvesting from simulation snapshots. (1) Blurring of the original image. (2) Binarization. (3) Dilation. (4) Extraction. See text for more details of the processing steps.}
\label{image-processing}
\end{figure}
\begin{enumerate}
\item Blur the snapshot image using blurring radius $r$ to generate a continuous density map.
\item Binarize the blurred image to clearly define the areas of interest.
\item Dilate each area of interest by radius $r$ to include less dense peripheral parts of the object.
\item Using the dilated area of interest as a mask, extract each object from the original snapshot.
\end{enumerate}
We used $r=150$, which was half of the maximum interaction range (300) of the Swarm Chemistry model. Note that this object harvesting method can extract not only small localized objects, but also large-scale global structures as well if they are made of spatially contiguous clusters of particles.

\subsection{Measurements}

\begin{figure}[t]
\centering
\includegraphics[width=\columnwidth]{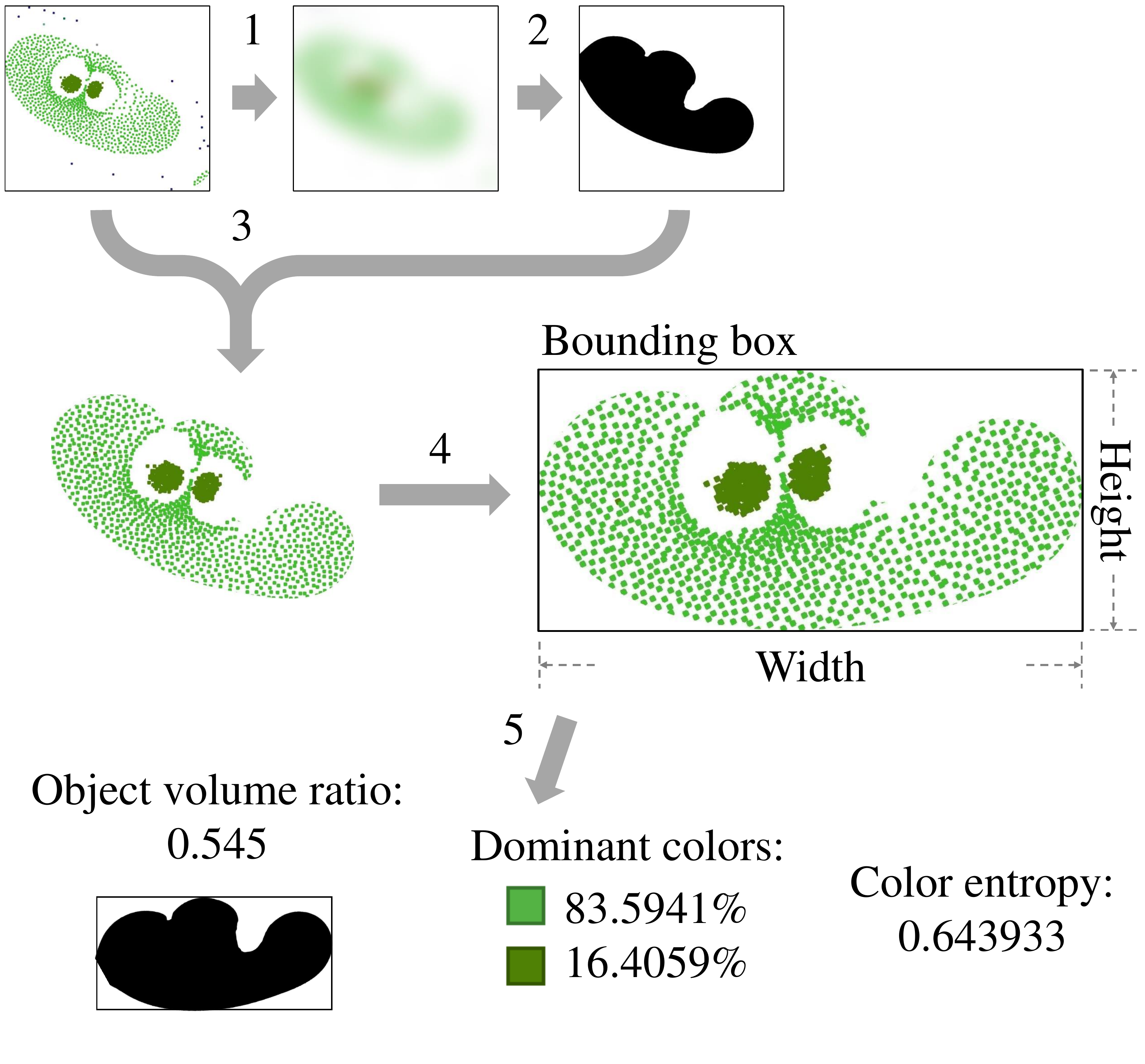}
\caption{Analysis of features of harvested objects. (1) Blurring of the harvested object. (2) Binarization. (3) Extraction. (4) Rotation. (5) Further measurements. See text for more details of the measurements.}
\label{pattern-analysis}
\end{figure}

The following measurements were calculated on the simulation snapshots and the harvested objects to characterize the evolutionary dynamics of each simulation run:
\begin{itemize}
\item On simulation snapshots \citep{sayama2011quantifying}
\begin{itemize}
\item Evolutionary exploration (EE):\\
Number of new colors that appeared at a specific time point for the first time within the simulation (roughly capturing how many novel particle types appeared)
\item Macroscopic structuredness (MS):\\
Kullback-Leibler divergence of an approximated pairwise particle distance distribution from that of a hypothetical case where particles were randomly and homogeneously spread over the space (capturing the extent to which particles formed nontrivial macroscopic structures)
\end{itemize}
\item On harvested objects (these were measured after the object was rotated appropriately to minimize the area of the bounding box; see Fig.~\ref{pattern-analysis})
\begin{itemize}
\item Position in space
\item Bounding box width and height (a longer side was always taken for width)
\item Bounding box height-width ratio
\item Bounding box area (= width $\times$ height)
\item Object volume ratio (= area of binarized object image / bounding box area)
\item Number of colors
\item Color entropy
\end{itemize}
\end{itemize}

All the image processing tasks involved in the automated object harvesting and object feature analysis were conducted using the image processing functions of Wolfram Research Mathematica 11.3.0.

\section{Results}

We ran six independent simulation runs for each of the ``short'' and ``long'' perturbation conditions (i.e, 12 runs total), each for 300,000 time steps. As a result, 3,600 snapshots were generated from these simulations. The automated object harvesting method was run on those snapshots, extracting 49,540 objects in total.

\subsection{Macroscopic trends of evolutionary activity}

\begin{figure}[t]
\includegraphics[width=\columnwidth]{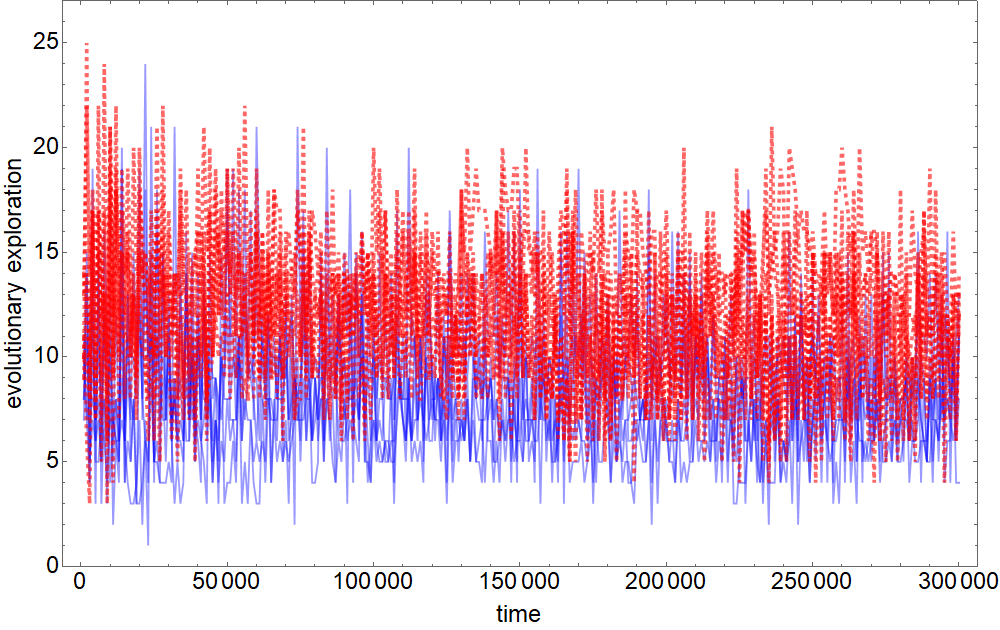}
\caption{Time series plots showing how the evolutionary exploration (EE) per snapshot changed over time in each simulation run. Blue (solid): Simulations with short perturbation conditions. Red (dashed): Simulations with long perturbation conditions.}
\label{macromeasures-EE}
\end{figure}

\begin{figure}[t]
\includegraphics[width=\columnwidth]{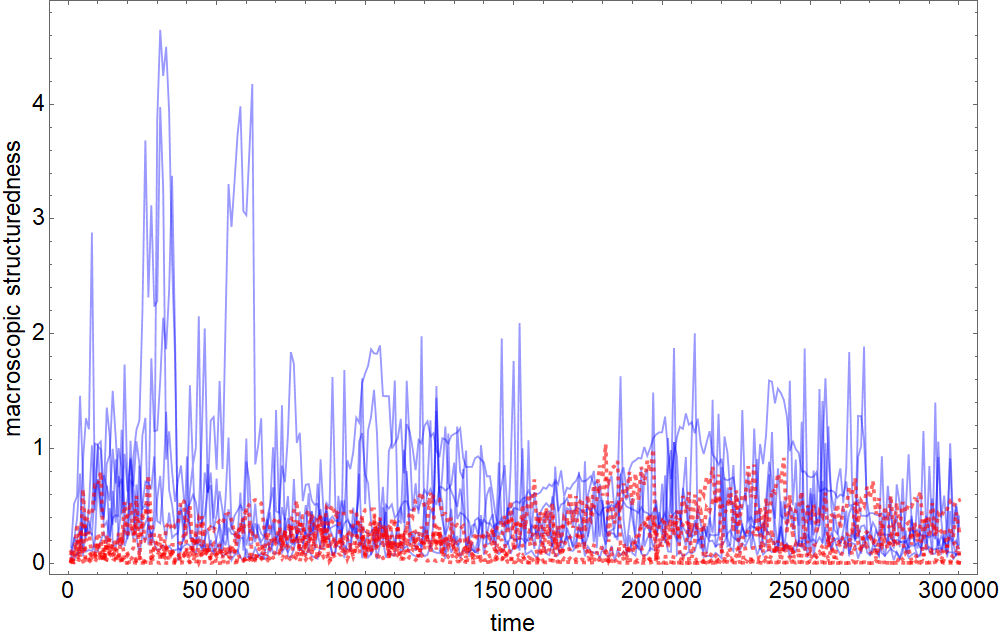}
\caption{Time series plots showing how the macroscopic structuredness (MS) per snapshot changed over time in each simulation run. Blue (solid): Simulations with short perturbation conditions. Red (dashed): Simulations with long perturbation conditions.}
\label{macromeasures-MS}
\end{figure}

Figures \ref{macromeasures-EE} and \ref{macromeasures-MS} show temporal changes of evolutionary exploration (EE) and macroscopic structuredness (MS) measurements, respectively, in simulations under the two perturbation conditions. It was observed in Fig.~\ref{macromeasures-EE} that EE dropped slightly after the initial 30,000--50,000 time steps, although production of new particle types was sustained at a moderate level for the entire duration of simulation. Fig.~\ref{macromeasures-MS} shows a significant difference in MS between the short and long perturbation conditions: the former maintained structured particle distributions while the latter failed to do so. This indicates that the too harsh environmental perturbation made it difficult for nontrivial macroscopic structures to be sustained.

\begin{figure*}[tbp]
\centering
\includegraphics[width=\textwidth]{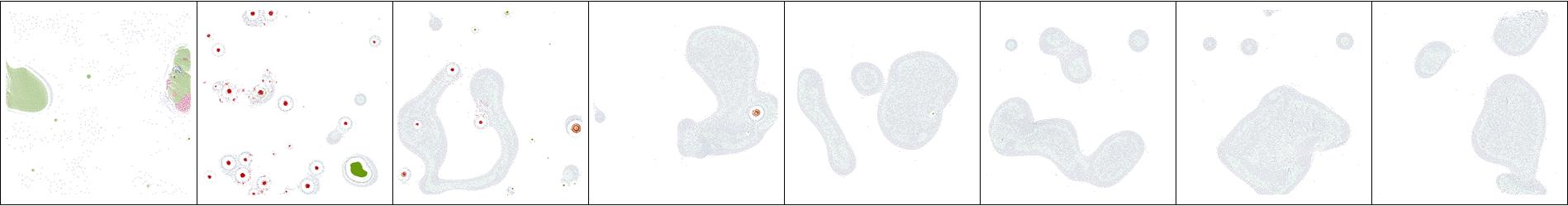}\\
~\\
\includegraphics[width=\textwidth]{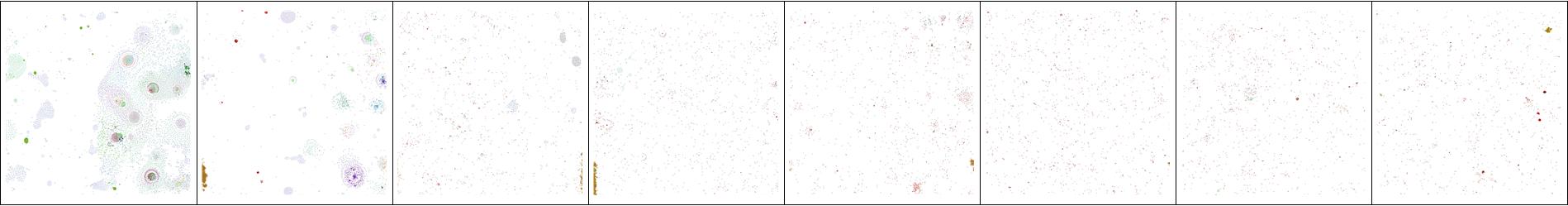}
\caption{Representative simulation runs that show typical evolutionary trends. Top: Under short perturbation conditions. Bottom: Under long perturbation conditions. Time flows from left to right ($t=10000,$ $30000,$ $50000,$ $100000,$ $150000,$ $200000,$ $250000,$ $300000$).}
\label{reprun}
\end{figure*}

Figure \ref{reprun} shows two representative simulation runs, one from the short perturbation condition and another from the long perturbation condition. As seen in the figure, a typical simulation run more or less settled down into an evolutionarily stable state after about 100,000 time steps, beyond which not much disruptive change occurred even though new particle types were continuously produced by environmental perturbations (Fig.~\ref{macromeasures-EE}). It was often observed that simulation runs under the short perturbation condition eventually ended up with a large-scale homogeneous swarming cloud (Fig.~\ref{reprun} top-right), while the runs under the long perturbation condition did not produce any such large-scale structure at the end (Fig.~\ref{reprun} bottom-right). Note that the earlier work on evolutionary Swarm Chemistry \citep{sayama2011seeking,sayama2011quantifying,sayama2012evolutionary} only looked at evolutionary dynamics up to 30,000 time steps due to limitations in computational resources, which corresponds to just the first two frames in Fig.~\ref{reprun}. Our new long-term simulations suggest that the evolutionary dynamics studied in the earlier work may have been just initial transient behaviors.

\subsection{Evolutionary trends of harvested objects}

\begin{figure}[t]
\includegraphics[width=\columnwidth]{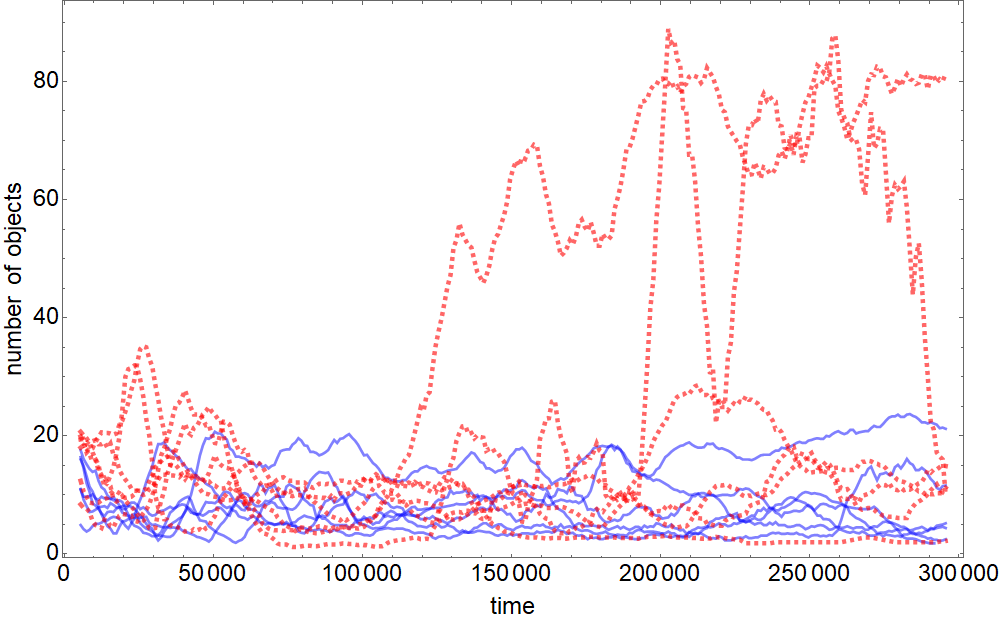}
\caption{Time series plots showing how the number of harvested objects per snapshot changed over time in each simulation run. Blue (solid): Simulations with short perturbation conditions. Red (dashed): Simulations with long perturbation conditions. Curves are smoothed by averaging over 10,000-step moving windows.}
\label{numplot}
\end{figure}

Figure \ref{numplot} shows the temporal changes of the number of harvested objects per snapshot in each simulation run. There was a slight upward trend for most of the runs, and in particular, the long perturbation condition produced a couple of extreme cases in which particles eventually evolved into many scattered tiny clusters. Once the system reached such a scattered, deserted state, it produced no more objects with nontrivial macroscopic structures or behaviors.

\begin{figure*}[p]
\centering
\begin{tabular}{cc}
Under short perturbation conditions & Under long perturbation conditions\\
\includegraphics[width=\columnwidth,height=1.3in]{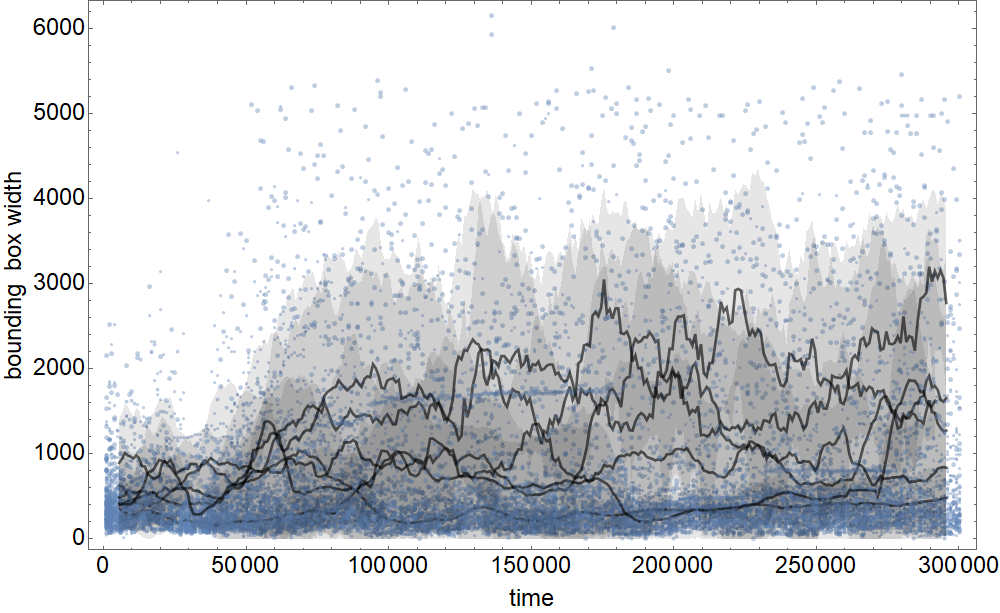} &
\includegraphics[width=\columnwidth,height=1.3in]{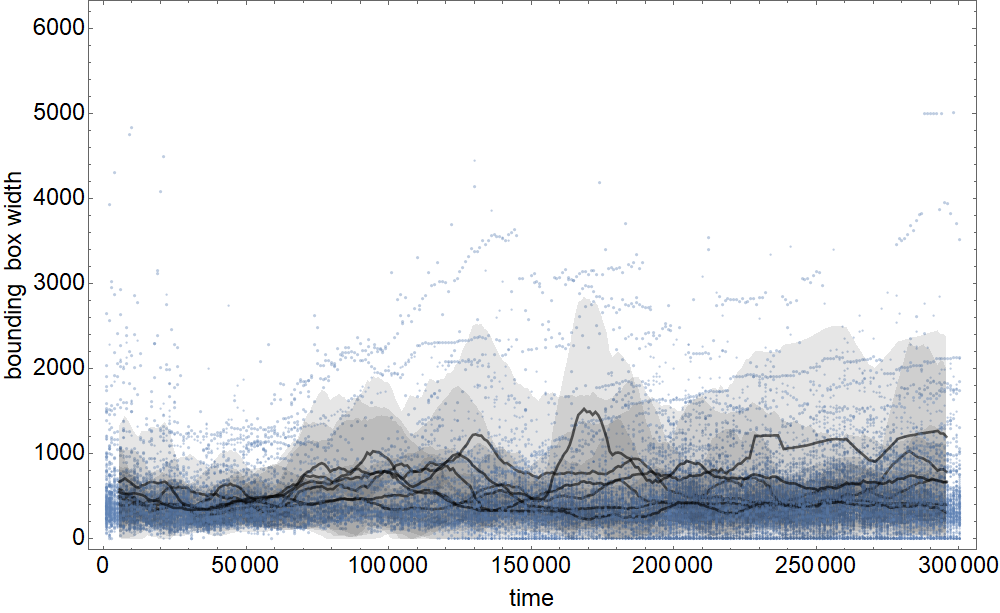} \\
\includegraphics[width=\columnwidth,height=1.3in]{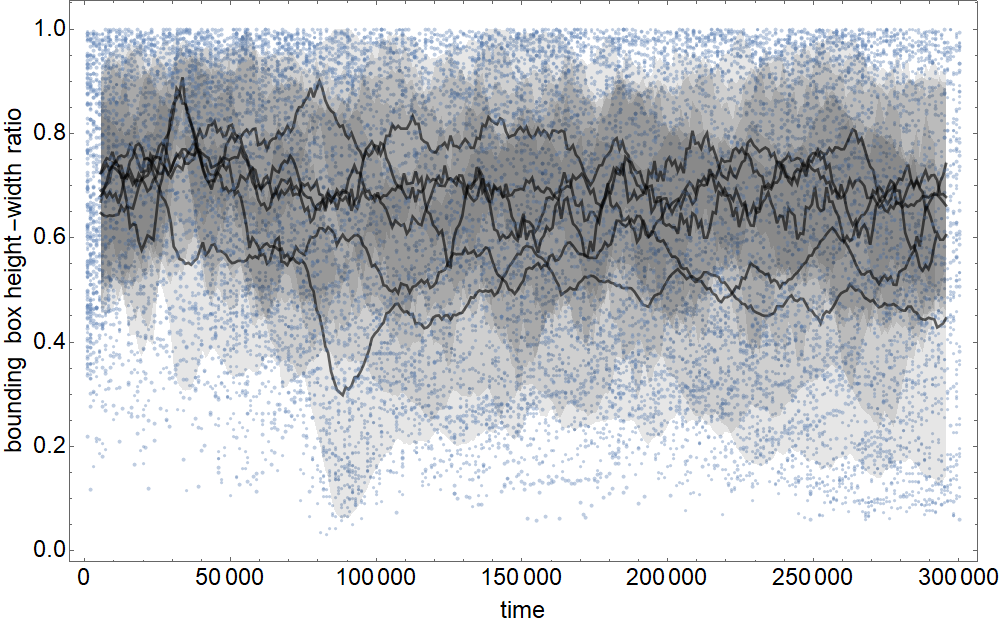} &
\includegraphics[width=\columnwidth,height=1.3in]{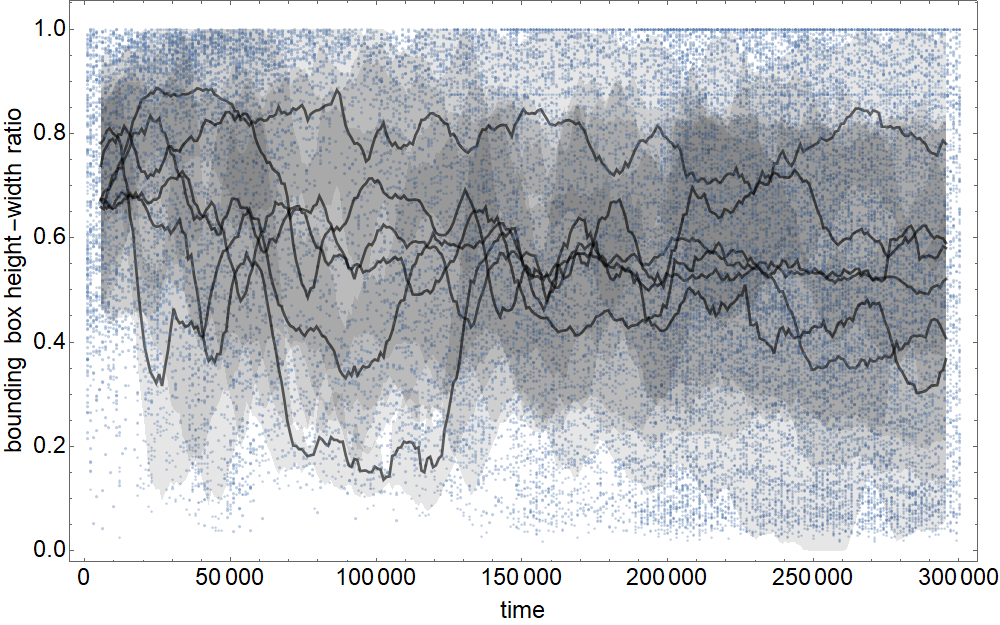} \\
\includegraphics[width=\columnwidth,height=1.3in]{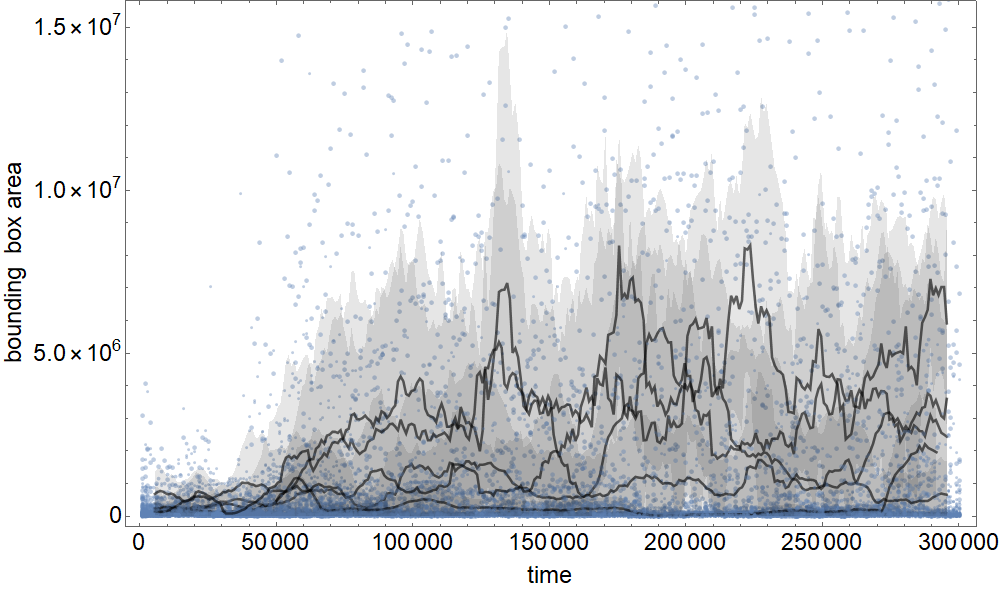} &
\includegraphics[width=\columnwidth,height=1.3in]{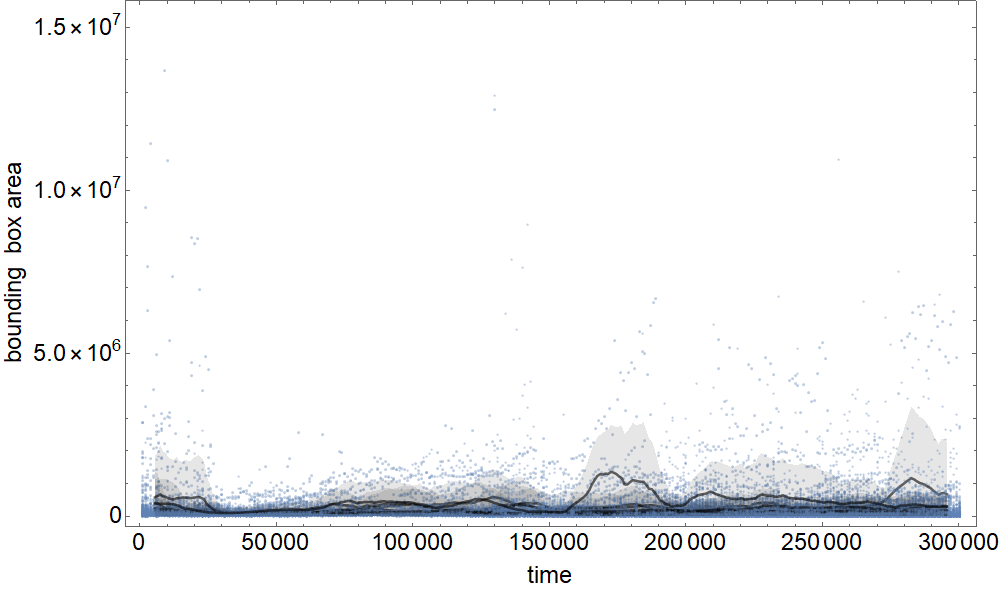} \\
\includegraphics[width=\columnwidth,height=1.3in]{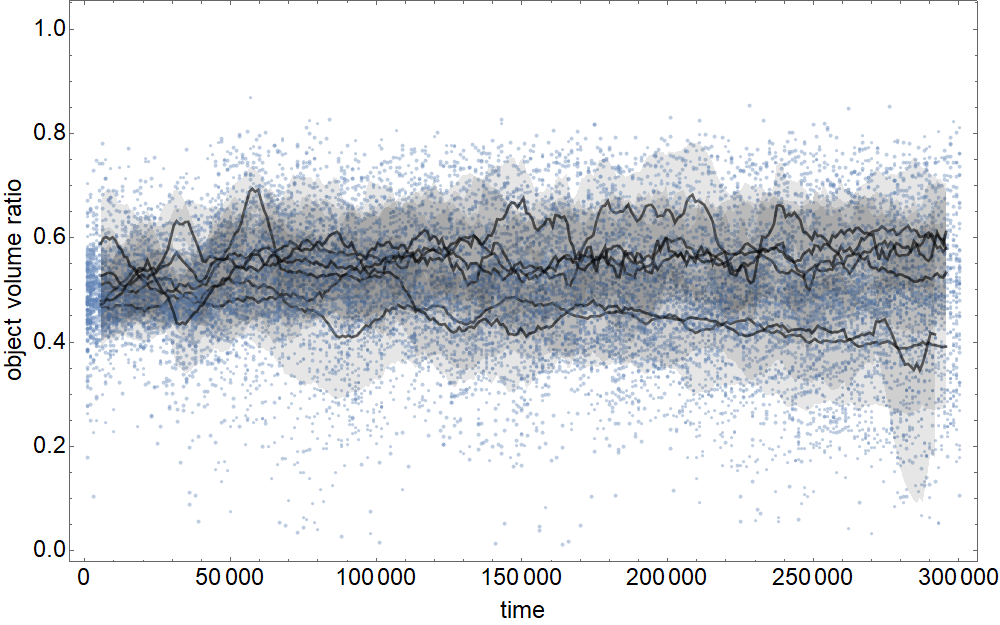} &
\includegraphics[width=\columnwidth,height=1.3in]{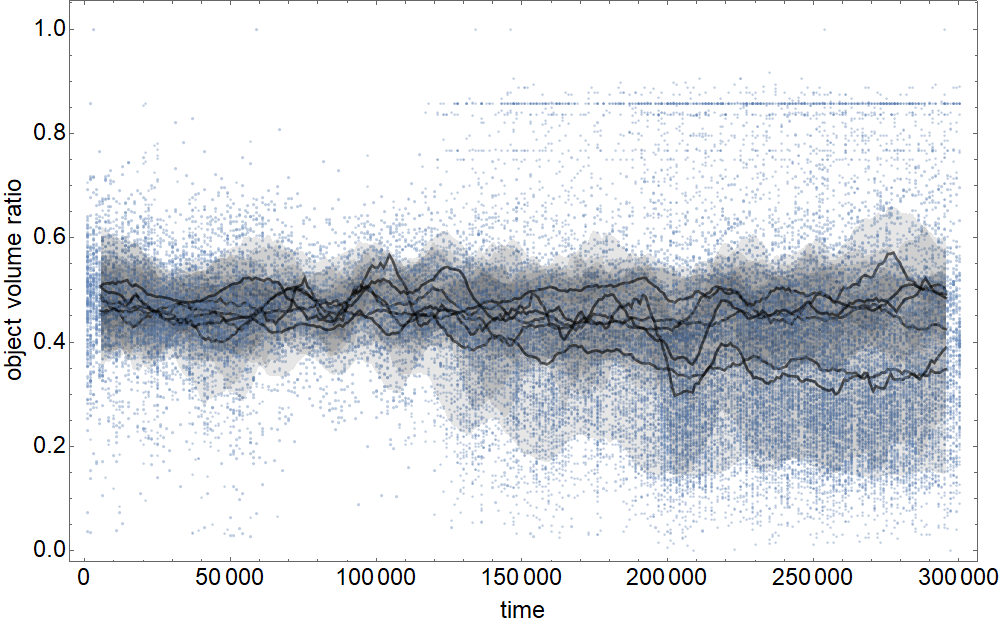} \\
\includegraphics[width=\columnwidth,height=1.3in]{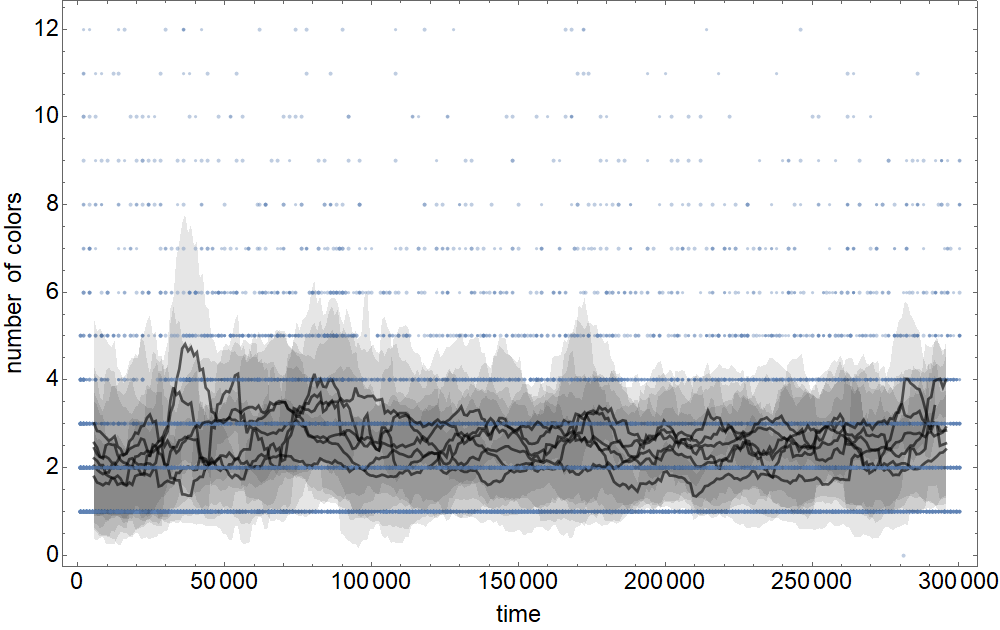} &
\includegraphics[width=\columnwidth,height=1.3in]{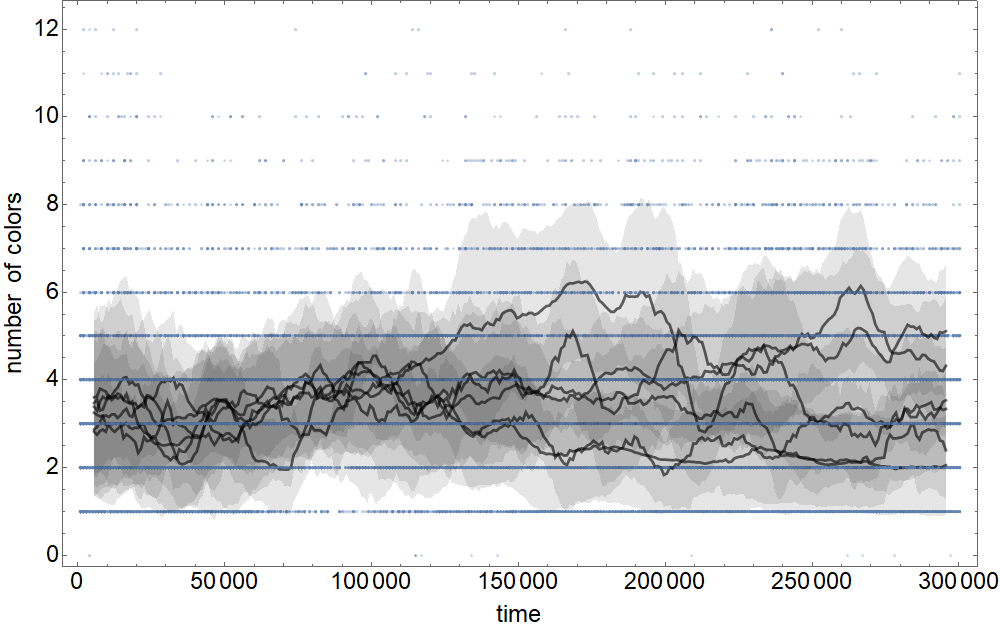} \\
\includegraphics[width=\columnwidth,height=1.3in]{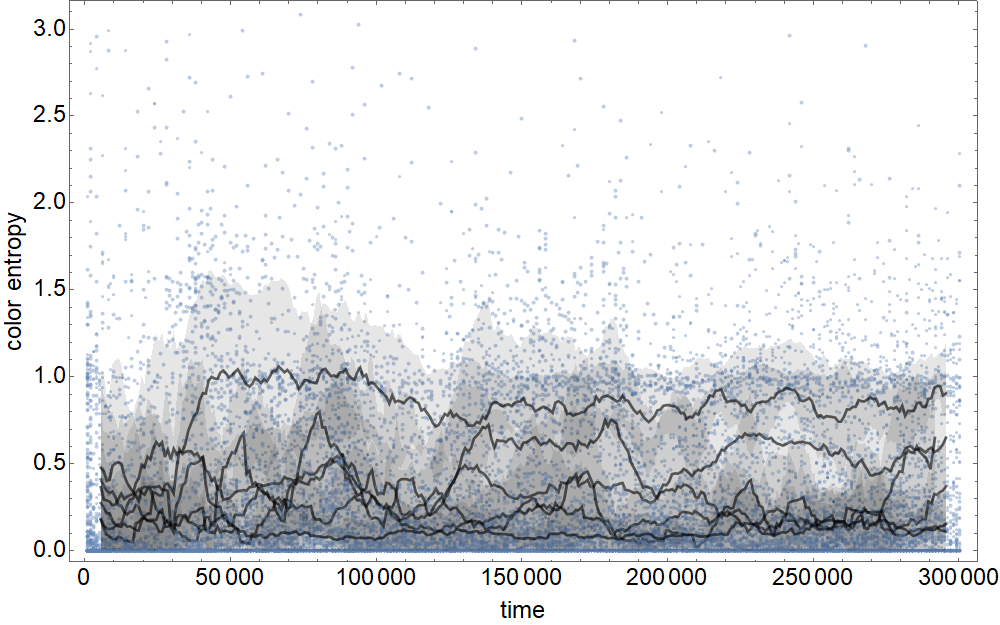} &
\includegraphics[width=\columnwidth,height=1.3in]{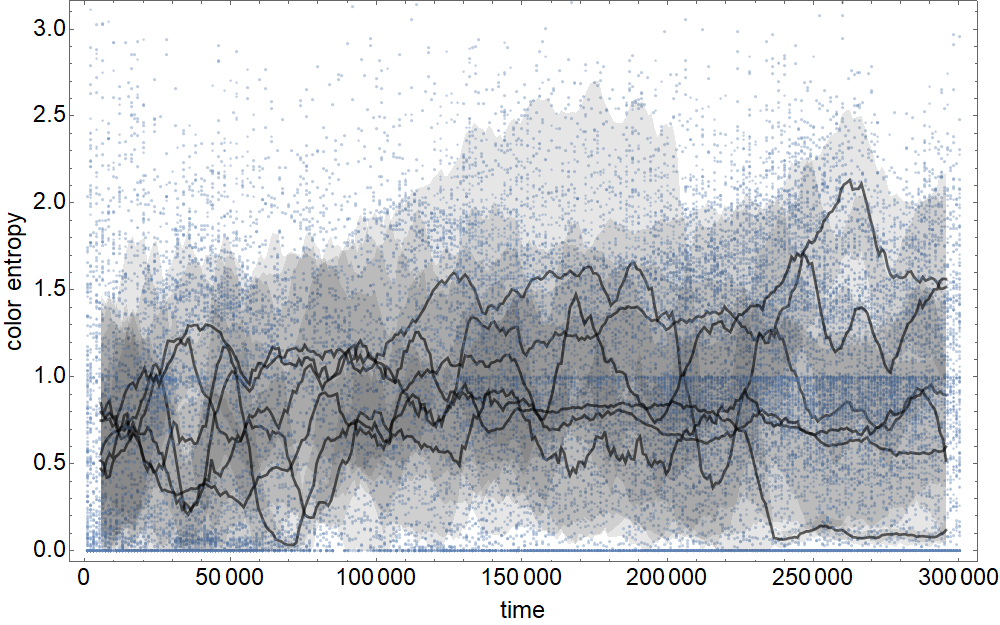}
\end{tabular}
\caption{Temporal changes of features of harvested objects. Black solid lines and gray shaded areas represent means and standard deviations, respectively. Curves are smoothed by averaging over 10,000-step moving windows. Blue dots show actual values coming from individual harvested objects. Left: Results under short perturbation conditions. Right: Results under long perturbation conditions.}
\label{measure-plots}
\end{figure*}

Figure \ref{measure-plots} shows how some of the features of harvested objects changed over time. In most of the measurements, there were clear long-term trends that were not fully visible within the initial 30,000 time steps that were the focus in the earlier work. For example, in the short perturbation cases (Fig.~\ref{measure-plots}, left), the bounding box width and area went up while the height-width ratio went down, which indicates that the particle population gradually became organized into a large-scale swarm with elongated shapes (see Fig.~\ref{reprun}, top). In the meantime, the number of colors and the color entropy did not show any significant changes, implying that the complexity of recipes regarding the number of kinetic types did not show long-term increase. Another important finding is the difference between the short and long perturbation conditions, most notably the much smaller size of objects (as seen in the bounding box width and area in Fig.~\ref{measure-plots}, right) that evolved under the long perturbation conditions. This quantitatively reflects the observation that particles tended to form much tinier clusters under such harsh environmental conditions (see Fig.~\ref{reprun}, bottom). It was also observed that the long perturbation conditions resulted in higher numbers of colors and higher color entropy than the short one (Fig.~\ref{measure-plots}, bottom two rows).

\subsection{Objects that evolved}

\begin{figure*}[tbp]
\centering
\begin{tabular}{c}
\includegraphics[width=0.75\textwidth]{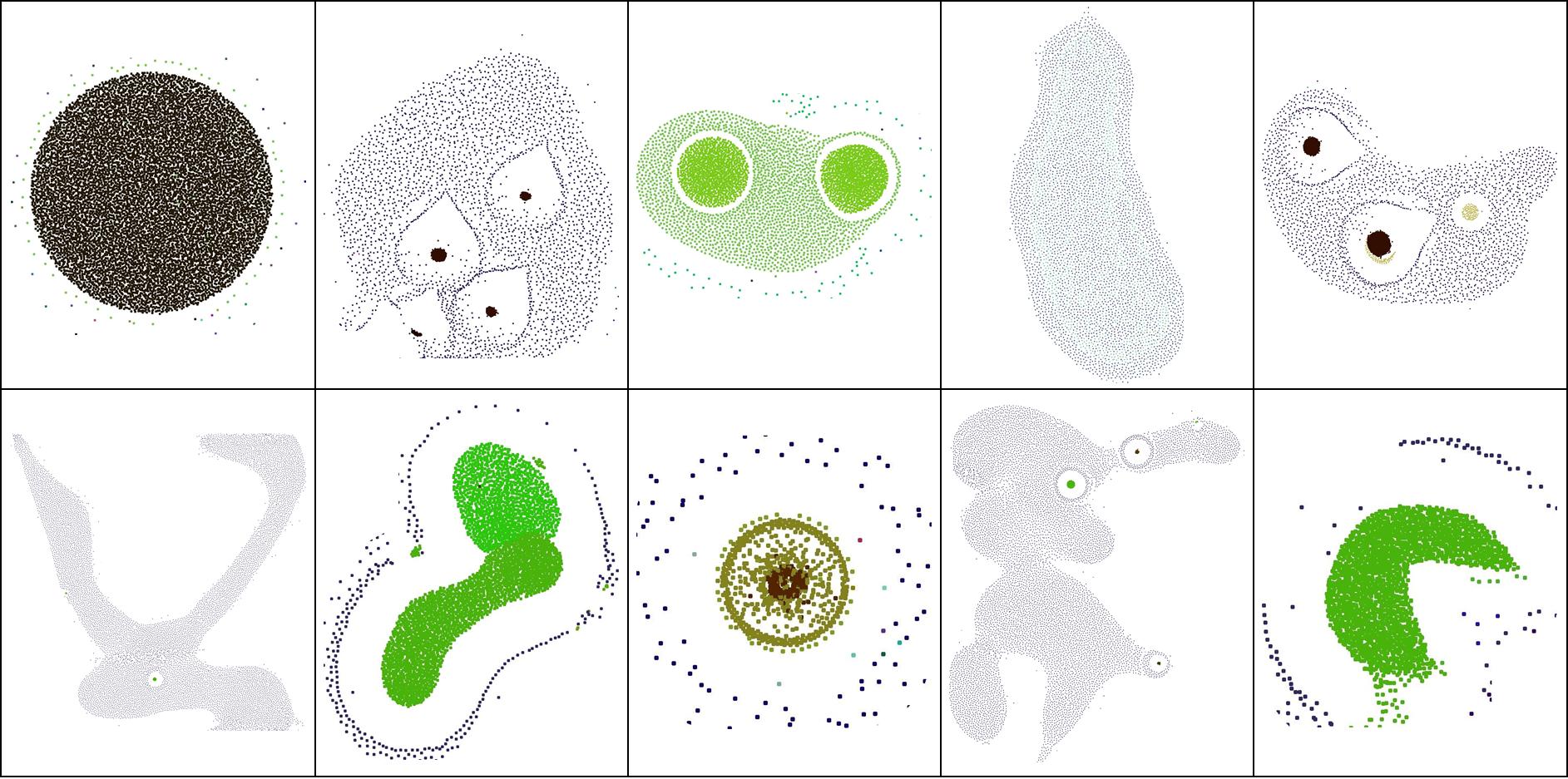}\\
 \\
\includegraphics[width=0.75\textwidth]{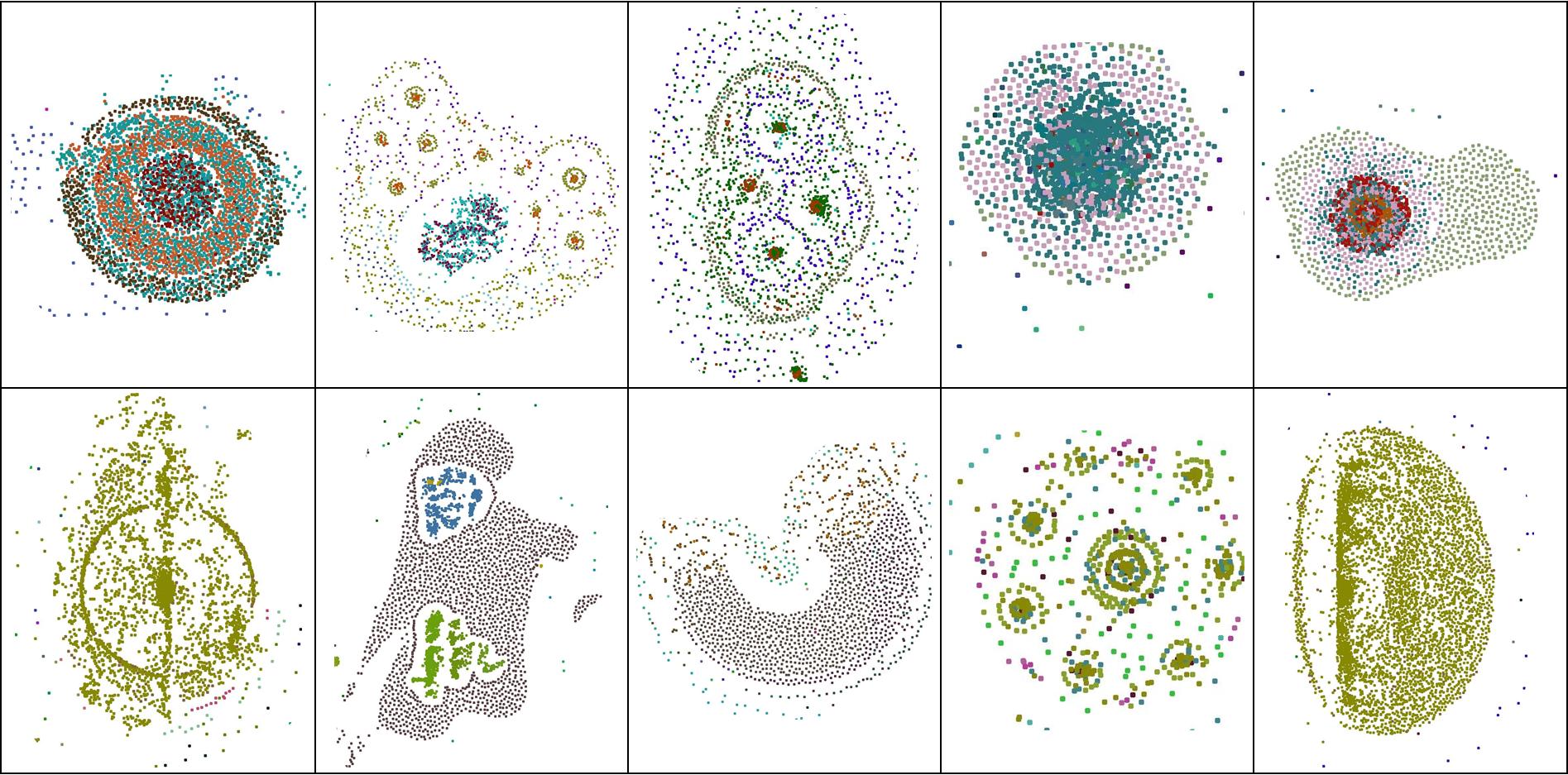}\\
\end{tabular}
\caption{Some fun examples subjectively selected from automatically harvested objects. Top: Objects generated under short perturbation conditions. Bottom: Objects generated under long perturbation conditions.}
\label{fun-patterns}
\end{figure*}

The results obtained so far may appear to collectively imply that the long-term evolutionary dynamics of the evolutionary Swarm Chemistry model would be rather limited in innovation and creativity. While this may be the case, the automated harvesting method, nonetheless, successfully captured a number of highly creative objects in the simulation runs. Figure \ref{fun-patterns} showcases some examples that were arbitrarily selected by the author based on subjective aesthetic criteria. Most of the objects shown in this figure were generated in the middle or later stages of long-term simulations, demonstrating the continuous autonomous creativity of evolutionary Swarm Chemistry. 

\begin{figure*}[tbp]
\centering
\begin{tabular}{c}
\includegraphics[width=0.9\textwidth]{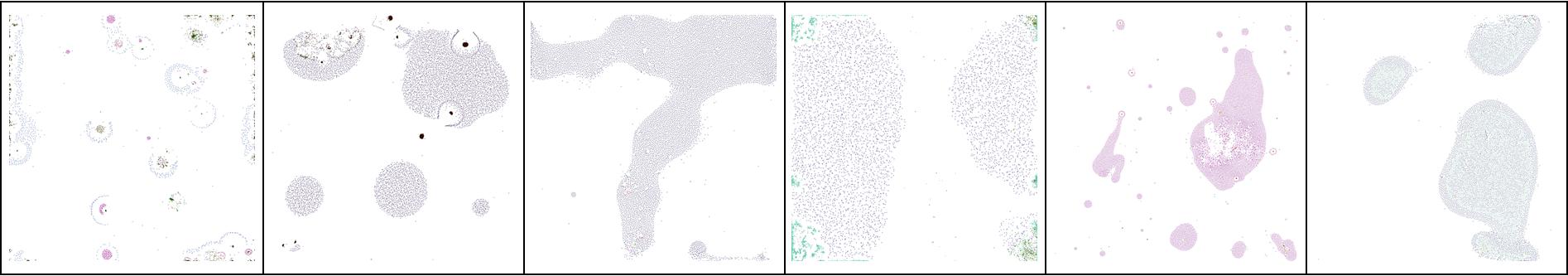}\\
 \\
\includegraphics[width=0.9\textwidth]{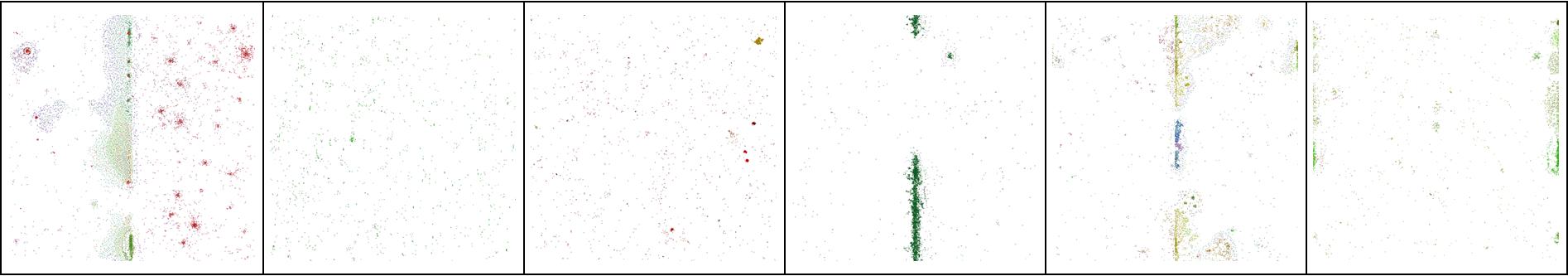}\\
\end{tabular}
\caption{Final states of simulations. Top: Final states under short perturbation conditions. Bottom: Final states under long perturbation conditions.}
\label{final-configs}
\end{figure*}

Figure \ref{fun-patterns} also shows that the objects generated under the short perturbation conditions tended to have well-defined, clearly identifiable structures, whereas those generated under the long perturbation conditions tended to have more scattered, fuzzy, noisy structures, often with a ``multi-cellular'' look. Such consistent differences of evolved objects between the two perturbation conditions imply that those differences should be understood as adaptive traits that helped objects survive different forms of perturbations.

Evidence of evolutionary adaptation can also be found in the final states of the simulations, which are visualized altogether in Fig.~\ref{final-configs}. Even though those simulation runs were completely independent of each other, the runs under the same perturbation condition generally evolved similar macroscopic outcomes, suggesting evolutionary convergence toward stable strategies that were effective in handling a particular type of environmental perturbations. Specifically, the runs under the short perturbation condition mostly produced large-sized homogeneous swarms (Fig.~\ref{final-configs}, top). In this condition, the perturbation (random change of local competition function) would last only 50 time steps, which would not allow for mutated recipes to spread too much (such a spread was actually captured in the second-to-right frame in Fig.~\ref{final-configs}, top). Given the large size of the swarm, such mutated recipes that might appear during a short period of perturbation would still be effectively contained and ``recruited back'' to the swarm. 

On the other hand, the same strategy did not evolve under the long perturbation condition where the perturbation would last as long as 500 time steps. In this harsh environment, some particles spontaneously ``figured out'' that one of the effective strategies would be to stay near the borders between left and right halves of the space without moving much so that, when a long tough time hits one half of the space, some of their peers can successfully maintain their recipes in the other half of the space nearby. This ``sit still at the border'' strategy can be seen as elongated clusters near the center and at the edges of the space in some of the results in Fig.~\ref{final-configs}, bottom. This is, in some sense, an artifact arising from a specific model assumption made in how environmental perturbations are induced, but it is still remarkable that evolution spontaneously discovered how to exploit it.

\section{Conclusions}

In this paper, we investigated the long-term dynamics of evolutionary Swarm Chemistry by conducting simulations that were ten times longer than in previous studies, under two different levels of environmental perturbations. Both macroscopic and microscopic structures were characterized; the latter was enabled by the newly developed automatic object harvesting method. The results indicated that, whereas certain evolutionary activities were sustained over an indefinitely long period of time, the evolutionary dynamics typically settled down after an initial transient period (to which the scope of earlier investigations was limited), and the evolutionary trends depended substantially on the extent of environmental perturbations. In most simulation runs, the system eventually fell into a seemingly evolutionarily stable state, and once this was reached there were little to no more disruptive evolutionary changes taking place. In this sense, it is still questionable whether evolutionary Swarm Chemistry has accomplished OEE. This observation, of course, depends on how one defines OEE \citep{taylor2016open}, as well as on the spatio-temporal scales of the analysis and many other model parameters. It is still possible that, if simulations were run within a much larger spatial domain for a much longer period of time, continuous evolutionary changes might be observed more significantly than the current study implies.

One thing that became empirically apparent in this study is the great difficulty of promoting continuous disruptive innovation in highly decentralized AChem-based evolutionary systems. We used the environmental perturbations to disrupt the \textit{status quo} established within the system, but natural selection was so powerful in finding sustainable stable strategies that could adapt to a ``meta-level'' environmental condition that subsumed those periodic perturbations. This observation leads us to the view that such convergence to a stable attractor state may be a general outcome of evolutionary systems, including real biological ones \citep{taylor2016open}. If we take this stance, evolutionary Swarm Chemistry can be considered biologically quite realistic.

Despite the difficulty of maintaining global-level evolutionary changes, the microscopic details of evolved objects, which were excavated at an unprecedented scale by the new automated harvesting method, successfully demonstrated the amazing autonomous creativity of evolutionary Swarm Chemistry. The current harvesting method simply processes saved bitmap images asynchronously, and therefore, it lacks the ability to pull out actual recipe information directly from the simulation. Our future plan includes implementing synchronous, online object harvesting that allows direct extraction of recipe information, which will greatly facilitate more detailed analysis of evolutionary dynamics as well as the application of evolutionary Swarm Chemistry as an autonomous creative engine. Such online recipe extraction may also allow for providing certain real-time feedback to the evolutionary process itself.

\section{Acknowledgments}

We thank the anonymous reviewers for their helpful comments that greatly helped improve the clarify of this paper. This material is based upon work supported by the National Science Foundation under Grant No. IIS-1319152.

\footnotesize
\bibliographystyle{apalike}
\bibliography{sayama}

\begin{thebibliography}{}

\bibitem[Banzhaf and Yamamoto, 2015]{banzhaf2015artificial}
Banzhaf, W. and Yamamoto, L. (2015).
\newblock {\em Artificial Chemistries}.
\newblock MIT Press.

\bibitem[Bedau and Brown, 1999]{bedau1999visualizing}
Bedau, M.~A. and Brown, C.~T. (1999).
\newblock Visualizing evolutionary activity of genotypes.
\newblock {\em Artificial Life}, 5(1):17--35.

\bibitem[Bedau et~al., 2000]{bedau2000open}
Bedau, M.~A., McCaskill, J.~S., Packard, N.~H., Rasmussen, S., Adami, C.,
  Green, D.~G., Ikegami, T., Kaneko, K., and Ray, T.~S. (2000).
\newblock Open problems in artificial life.
\newblock {\em Artificial Life}, 6(4):363--376.

\bibitem[Bedau and Packard, 1992]{bedau1992measurement}
Bedau, M.~A. and Packard, N.~H. (1992).
\newblock Measurement of evolutionary activity, teleology, and life.
\newblock In {\em Artificial Life II}, pages 431--461. Addison-Wesley.

\bibitem[Bedau et~al., 1998]{bedau1998classification}
Bedau, M.~A., Snyder, E., and Packard, N.~H. (1998).
\newblock A classification of long-term evolutionary dynamics.
\newblock In {\em Artificial Life VI}, pages 228--237. MIT Press.

\bibitem[Dittrich et~al., 2001]{dittrich2001artificial}
Dittrich, P., Ziegler, J., and Banzhaf, W. (2001).
\newblock Artificial chemistries---a review.
\newblock {\em Artificial Life}, 7(3):225--275.

\bibitem[Reynolds, 1987]{reynolds1987}
Reynolds, C.~W. (1987).
\newblock Flocks, herds and schools: A distributed behavioral model.
\newblock {\em Computer Graphics}, 21(4):25--34.

\bibitem[Sayama, 2009]{sayama2009swarm}
Sayama, H. (2009).
\newblock Swarm chemistry.
\newblock {\em Artificial Life}, 15(1):105--114.

\bibitem[Sayama, 2011]{sayama2011seeking}
Sayama, H. (2011).
\newblock Seeking open-ended evolution in swarm chemistry.
\newblock In {\em Proceedings of the 2011 IEEE Symposium on Artificial Life
  (IEEE ALIFE 2011)}, pages 186--193. IEEE.

\bibitem[Sayama, 2012]{sayama2012evolutionary}
Sayama, H. (2012).
\newblock Evolutionary swarm chemistry in three-dimensions.
\newblock In {\em Artificial Life 13: Proceedings of the Thirteenth
  International Conference on the Simulation and Synthesis of Living Systems},
  pages 576--577. MIT Press.

\bibitem[Sayama, 2018]{sayama2018complexity}
Sayama, H. (2018).
\newblock Complexity, development, and evolution in morphogenetic collective
  systems.
\newblock {\em arXiv preprint arXiv:1801.02086}.

\bibitem[Sayama and Wong, 2011]{sayama2011quantifying}
Sayama, H. and Wong, C. (2011).
\newblock Quantifying evolutionary dynamics of swarm chemistry.
\newblock In {\em Proceedings of ECAL 2011}, pages 729--730. MIT Press.

\bibitem[Schmickl et~al., 2016]{schmickl2016life}
Schmickl, T., Stefanec, M., and Crailsheim, K. (2016).
\newblock How a life-like system emerges from a simple particle motion law.
\newblock {\em Scientific Reports}, 6:37969.

\bibitem[Soros and Stanley, 2014]{soros2014identifying}
Soros, L.~B. and Stanley, K.~O. (2014).
\newblock Identifying necessary conditions for open-ended evolution through the
  artificial life world of chromaria.
\newblock In {\em Artificial Life 14: Proceedings of the Fourteenth
  International Conference on the Synthesis and Simulation of Living Systems},
  pages 793--800. MIT Press.

\bibitem[Stanley et~al., 2017]{stanley2017open}
Stanley, K.~O., Lehman, J., and Soros, L. (2017).
\newblock Open-endedness: The last grand challenge you've never heard of.
\newblock
  \url{https://www.oreilly.com/ideas/open-endedness-the-last-grand-challenge-youve-never-heard-of}.
\newblock Accessed on April 1, 2018.

\bibitem[Taylor et~al., 2016]{taylor2016open}
Taylor, T., Bedau, M., Channon, A., Ackley, D., Banzhaf, W., Beslon, G.,
  Dolson, E., Froese, T., Hickinbotham, S., Ikegami, T., et~al. (2016).
\newblock Open-ended evolution: perspectives from the oee workshop in york.
\newblock {\em Artificial Life}, 22(3):408--423.

\end{thebibliography}

\end{document}